\documentclass[aps,prl,twocolumn,showpacs,superscriptaddress,amsmath,floatfix]{revtex4-1}
\usepackage{graphicx}
\usepackage{bm}

\begin{document}

\title{Impact of Active-Sterile Neutrino Mixing on Supernova Explosion and Nucleosynthesis}

\author{Meng-Ru Wu}
\affiliation{Institut f{\"u}r Kernphysik (Theoriezentrum), Technische
  Universit{\"a}t Darmstadt, Schlossgartenstra{\ss}e 2, 64289
  Darmstadt, Germany} 
\author{Tobias Fischer}
\affiliation{Institute for Theoretical Physics, University of Wroc{\l}aw,
pl. M. Borna 9, 50-204 Wroc{\l}aw, Poland} 
\author{Lutz Huther}
\affiliation{Institut f{\"u}r Kernphysik (Theoriezentrum), Technische
  Universit{\"a}t Darmstadt, Schlossgartenstra{\ss}e 2, 64289
  Darmstadt, Germany} 
\author{Gabriel Mart{\'i}nez-Pinedo}
\affiliation{Institut f{\"u}r Kernphysik (Theoriezentrum), Technische
  Universit{\"a}t Darmstadt, Schlossgartenstra{\ss}e 2, 64289
  Darmstadt, Germany} 
\affiliation{GSI Helmholtzzentrum f\"ur Schwerioneneforschung,
  Planckstra{\ss}e~1, 64291 Darmstadt, Germany} 
\author{Yong-Zhong Qian}
\affiliation{School of Physics and Astronomy, University of Minnesota,
  Minneapolis, MN 55455, USA} 

\date{\today}

\begin{abstract}
We show that for the active-sterile flavor mixing parameters suggested by the
reactor neutrino anomaly, substantial $\nu_e$-$\nu_s$ and 
$\bar\nu_e$-$\bar\nu_s$ conversion occurs in regions with electron fraction
of $\approx 1/3$ near the core of
an $8.8\,M_\odot$ electron-capture supernova. 
Compared to the case without such conversion, 
the neutron-richness of the ejected material is enhanced to allow
production of elements from Sr, Y, and Zr up to Cd in broad agreement with 
observations of the metal-poor star HD~122563.
Active-sterile flavor conversion also strongly suppresses neutrino heating 
at times when it is important for 
the revival of the shock. Our results suggest that simulations of supernova 
explosion and nucleosynthesis may be used to constrain active-sterile 
mixing parameters in combination
with neutrino experiments and cosmological considerations.
\end{abstract}


\maketitle

The number of active neutrino flavors that participate in weak interactions 
in the Standard Model has been precisely determined from the $Z^0$
decay width to be $2.984\pm 0.008$ \cite{ALEPH:2005ab}. 
Nearly all the vacuum mixing parameters for active neutrinos have been 
measured, except for the mass hierarchy and the CP violating phase
\cite{Beringer:1900zz}. However, recent results on oscillations of
neutrinos from reactors, radioactive sources, and accelerators indicate that
the standard mixing scenario involving only the three active flavors may
not be complete 
\cite{Aguilar:2001ty,AguilarArevalo:2010wv,Aguilar-Arevalo:2013pmq,Mention:2011rk,Acero:2007su}. 
Light sterile neutrinos ($\nu_s$) of the eV mass scale that mix with active neutrinos
have been proposed as the simplest extension beyond the standard scenario
to explain these anomalies (see \cite{Abazajian:2012ys} for a recent review). 
The effective 
relativistic degree of freedom inferred from both Cosmic Microwave 
Background data and Big Bang Nucleosynthesis studies has large enough uncertainties
to also allow such sterile neutrinos to exist
\citep{Archidiacono:2013xxa,Hamann:2010bk,Mangano:2011ar}.
There is some tension between the sterile neutrino parameters obtained from
different oscillation experiments and cosmological constraints \citep{Kopp:2013vaa,Giunti:2013aea,Mirizzi:2013kva}. 
This will be resolved by
future experiments including solar neutrino measurements \citep{Abazajian:2012ys}.

In addition to important implications for experiments and cosmology,
mixing of sterile with active neutrinos may lead to multiple
Mikheyev-Smirnov-Wolfenstein (MSW) resonances in matter
\cite{Wolfenstein,MS,Kainulainen:1990bn}, which could have interesting effects 
in supernovae \cite{Nunokawa:1997ct,Kainulainen:1990bn,Shi:1993ee,Pastor:1994nx,%
Caldwell:1999zk,Fetter:2002xx,Beun:2006ka,Hidaka:2006sg,Choubey:2007ga,%
Hidaka:2007se,Fuller:2009zz,Tamborra:2011is}. 
In particular, \cite{Nunokawa:1997ct} pointed out that for light sterile neutrinos of 
1--100~eV, $\nu_e$-$\nu_s$ and $\bar\nu_e$-$\bar\nu_s$
conversion could have significant impact on supernova explosion and nucleosynthesis. 
As such conversion changes the electron fraction $Y_e$,
and hence the matter potential determining the MSW resonances, this feedback
should be included in a full treatment of the problem. Possible feedback effects 
were noted in \cite{Nunokawa:1997ct} and taken into account in some later works
(e.g., \cite{Fetter:2002xx,Beun:2006ka,Tamborra:2011is}), which
mostly focused on the outer resonances at baryon densities of
$\rho<10^8$~g~cm$^{-3}$. 
 
In this Letter we examine active-sterile flavor conversion (ASFC)
in the region of the inner resonances (IR) where $\rho\sim 10^9$--$10^{12}$~g~cm$^{-3}$ 
and evaluate its impact on supernova dynamics and nucleosynthesis.
A supernova starts 
when a massive star undergoes gravitational core collapse
at the end of its life. Upon reaching
supra-nuclear density, the core bounces to launch a shock and
a protoneutron star forms and cools by emitting all three flavors of active (anti)neutrinos.
As the shock propagates out of the core, it loses energy by dissociating nuclei
in the material falling through it into free nucleons.
At the same time, the dissociated material behind the shock gains energy from
neutrino heating mainly by the following reactions:
\begin{subequations}
\begin{eqnarray}
\nu_e+n&\to&p+e^-,\label{eq-nuen}\\
\bar\nu_e+p&\to&n+e^+.\label{eq-anuep}
\end{eqnarray}
\end{subequations}
This is the essence of the so-called delayed neutrino-heating explosion mechanism \cite{Bethe:1984ux},
which has been shown to result in explosions in recent supernova simulations
\cite{Mueller:2012is,Suwa:2012xd,Bruenn:2012mj}. 
Reactions~(\ref{eq-nuen}) and (\ref{eq-anuep}) and their reverse reactions
are essential to determining the $Y_e$ and hence, nucleosynthesis in any 
neutrino-heated ejecta \cite{Qian:1993dg,Qian:1996xt,Arcones:2012wj}.
ASFC of the $\nu_e$-$\nu_s$ and $\bar\nu_e$-$\bar\nu_s$ types 
may influence the rates of reactions~(\ref{eq-nuen}) and (\ref{eq-anuep}), 
and consequently, affect supernova dynamics and nucleosynthesis.

To treat ASFC in supernovae, we assume 
effective 2-$\nu$ mixing between $\nu_e$ and $\nu_s$ ($\bar\nu_e$ and $\bar\nu_s$). 
This is justified as the magnitude of the mass-squared difference 
$\delta m^2\sim{\cal{O}}(\pm1)\,{\rm eV}^2$
between the relevant vacuum mass eigenstates greatly exceeds those for mixing
among active neutrinos, and collective oscillations among active neutrinos are expected 
to be suppressed due to the high matter density in the region of interest to us
\cite{EstebanPretel:2008ni}. We further assume that nearly all neutrinos 
freely stream through this region due to their 
large mean-free-path.
Under these assumptions, a $\nu_e$-$\nu_s$ resonance occurs when
\begin{equation}\label{eq-res}
\frac{\delta m^2}{2E_\nu}\cos{2\theta}=
\frac{3\sqrt{2}}{2}G_F n_b\left(Y_e-\frac{1}{3}\right)=V_{\nu_e}^{\rm eff},
\end{equation}
where $\theta$ is the vacuum mixing 
angle, $E_\nu$ is the neutrino energy (with average values of 
$\langle E_\nu\rangle\sim 10$--15~MeV), 
$G_F$ is the Fermi coupling constant, and
$n_b=\rho/m_u$ is the baryon number density with $m_u$ being the atomic mass unit.
The right-hand side of Eq.~(\ref{eq-res}) is the effective potential $V_{\nu_e}^{\rm eff}$
from $\nu_e$ forward scattering on neutrons, protons, and $e^\pm$ in matter
\cite{Kainulainen:1990bn,Nunokawa:1997ct}. We neglect the contribution of
$\nu$-$\nu$ forward scattering \cite{Sigl:1992fn,Fuller:1987,Pantaleone:1992eq},
which is only $\sim 1\%$ of $V_{\nu_e}^{\rm eff}$ around the IR region.

\begin{figure}
\includegraphics[angle=-90,width=\linewidth]{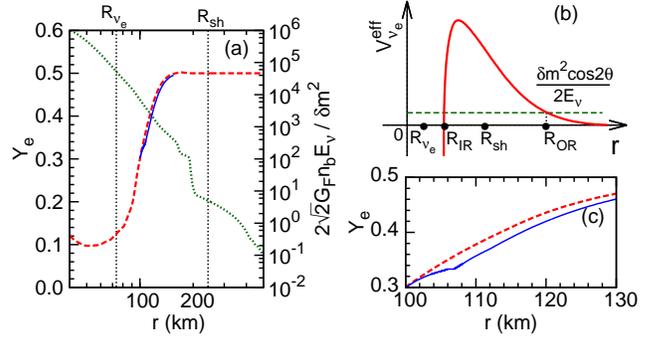}
\caption{
(Color online) (a) $Y_e$ profile with (blue solid curve) and 
without (red dashed curve) ASFC for
$\delta m^2=1.75$~eV$^2$ and $\sin^2{2\theta}=0.10$, 
and $2\sqrt{2}G_Fn_bE_\nu/\delta m^2$ for $E_\nu=10$~MeV (green dotted curve) 
at $t_{\rm pb}\sim 34$~ms. (b) Schematic plot of $V^{\rm eff}_{\nu_e}$ (red solid curve) 
as a function of radius, illustrating the positions of different resonances with
respect to the $\nu_e$-sphere ($R_{\nu_e}$) and the shock ($R_{\rm sh}$).
(c) Blow-up of the $Y_e$ profiles in (a) for $r\gtrsim R_{\rm IR}$.
\label{fig-profile}}
\end{figure}

In supernovae, the central $Y_e$ is $\lesssim 0.3$ after core bounce
as a result of electron capture on nuclei and free protons during core collapse
\cite{Hix:2003fg,Langanke:2003ii}. With increasing radius, 
$Y_e$ becomes as low as 0.1 near the neutrinospheres,
due to electron capture on shock-dissociated material.
At even larger radii, $Y_e$ increases to $\sim 0.5$ due to reactions 
(\ref{eq-nuen}) and (\ref{eq-anuep}) (see Fig.~\ref{fig-profile}a). 
Consequently, above the $\nu_e$-sphere at $r=R_{\nu_e}$, there is a radius, $R_\text{IR}$,
where $Y_e= 1/3^+$ but $\sqrt{2}G_Fn_b\gg\delta m^2/2E_\nu$, and
Eq.~(\ref{eq-res}) is satisfied for $\delta m^2 >0$ (normal hierarchy). 
For typical density profiles, this IR corresponds to
$\rho\sim 10^{9}$--$10^{12}$~g~cm$^{-3}$.
A second outer resonance (OR) occurs
further out at larger values of $Y_e$ once $n_b$ drops enough to satisfy Eq.~(\ref{eq-res})
(see Fig.~\ref{fig-profile}b). 
The condition for a $\bar\nu_e$-$\bar\nu_s$ resonance 
differs from Eq.~(\ref{eq-res}) by an opposite sign of the effective potential:
\begin{equation}\label{eq-resb}
\frac{\delta m^2}{2E_\nu}\cos{2\theta}=
-\frac{3\sqrt{2}}{2}G_F n_b\left(Y_e-\frac{1}{3}\right)=V_{\bar\nu_e}^{\rm eff}.
\end{equation}
If $\delta m^2>0$, an IR occurs for $\bar\nu_e$-$\bar\nu_s$ conversion
for $Y_e= 1/3^-$ but there is no
OR in this case. If $\delta m^2<0$ (inverted hierarchy),
there would be only an IR for $\nu_e$-$\nu_s$ conversion but both an IR and 
an OR for $\bar\nu_e$-$\bar\nu_s$ conversion.

It is clear from the above discussion that independent of the mass hierarchy
and for $|\delta m^2|\lesssim 10$~eV$^2$, there is always an IR for both $\nu_e$-$\nu_s$ 
and $\bar\nu_e$-$\bar\nu_s$ conversion at $Y_e\approx 1/3$ in supernovae. 
[There is only an OR for $\nu_x$-$\nu_s$ 
or $\bar\nu_x$-$\bar\nu_s$ ($x=\mu$, $\tau$) conversion as
$V^{\rm eff}_{\nu_x}\propto (1-Y_e)$]. However, an inverted hierarchy appears 
disfavored by neutrino mass constraints from the Cosmic Microwave Background and 
Tritium decay experiments \cite{Hinshaw:2012fq,Ade:2013lta,Kraus:2004zw}. 
Therefore, we will not discuss this case further but will focus on the IR 
at $Y_e\approx 1/3$ for a normal hierarchy with $\delta m^2>0$ below. 

As $\sqrt{2}G_Fn_b\gg\delta m^2/2E_\nu$ in the region of interest, 
the local effective neutrino mass eigenstates are determined by 
$V_{\nu_e}^{\rm eff}$ except near the IR. 
At $r<R_{\rm IR}$ where $Y_e<1/3$ gives $V_{\nu_e}^{\rm eff}<0$, 
the lighter (heavier) effective mass eigenstate $\nu_L$ ($\nu_H$)
is approximately $\nu_e$ ($\nu_s$), while
at $r>R_{\rm IR}$ where $Y_e>1/3$ gives $V_{\nu_e}^{\rm eff}>0$, 
$\nu_L$ ($\nu_H$) is approximately $\nu_s$ ($\nu_e$). 
With $V_{\bar\nu_e}^{\rm eff}=-V_{\nu_e}^{\rm eff}$, the situation 
is exactly the opposite for the effective antineutrino mass eigenstates.
As $\nu_e$ ($\bar\nu_e$) cross $R_{\rm IR}$ from below,
the probability that they hop from $\nu_L$ ($\bar\nu_H$) to 
$\nu_H$ ($\bar\nu_L$) can be approximated by the Landau-Zener 
formula \cite{Landau:1932,Zener:1932ws}:
\begin{equation}\label{eq-adia}
P_{\rm hop}(E_\nu,\mu)=\exp\left(-\frac{\pi^2}{2}\frac{\delta r}{L}\right),
\end{equation}
where 
$\delta r=\delta m^2\sin 2\theta/(\mu E_\nu
|dV_{\nu_e(\bar\nu_e)}^{\rm eff}/dr|_{\rm res})$ 
is the width of the resonance region with $\mu$ the cosine of the angle 
between the $\nu_e$ ($\bar\nu_e$) momentum and the radial direction,
and $L=4\pi E_\nu/(\delta m^2\sin 2\theta)$ is the oscillation 
length at resonance \cite{Robertson:2012ib}.
Taking $\delta m^2=1.75$~eV$^2$ 
and $\sin^2{2\theta}=0.10$, which are the best-fit parameters inferred from 
reactor neutrino experiments \cite{Kopp:2013vaa}, we
obtain $L\approx 45$~m for $E_\nu=10$~MeV. As $|Y_e-1/3|\ll 1$ at the IR, 
$|dV_{\nu_e(\bar\nu_e)}^{\rm eff}/dr|_{\rm res}\approx(3\sqrt{2}/2)G_F
(\rho_{\rm res}/m_u)|dY_e/dr|_{\rm res}$. Using the same mixing parameters 
as above and taking $\rho_{\rm res}=10^9$~g~cm$^{-3}$, 
$|dY_e/dr|_{\rm res}=10^{-2}$~km$^{-1}$, 
and $\mu=1$, we obtain $\delta r\approx 48$~m for $E_\nu=10$~MeV. 
In the above example, $\delta r\gtrsim L$ gives $P_{\rm hop}\sim 0$, which
means that $\nu_e$ ($\bar\nu_e$) produced originally as $\nu_L$ ($\bar\nu_H$) 
will stay in their local effective mass eigenstates, and after crossing the IR, 
be completely converted into $\nu_s$ ($\bar\nu_s$). 

In general, $\delta r/L\propto(\delta m^2\sin2\theta/E_\nu)^2
(\rho|dY_e/dr|)^{-1}_{\rm res}/\mu$ and therefore,
$P_{\rm hop}(E_\nu,\mu)$ is sensitive to
$dY_e/dr$ at $Y_e\approx 1/3$. We define $E_{0.5}$ as the 
$E_\nu$ corresponding to $P_{\rm hop}=0.5$ for $\mu=1$.
It is clear from Eq.~(\ref{eq-adia}) and the above discussion
that most of the $\nu_e$ ($\bar\nu_e$)
with $E_\nu<E_{0.5}$ will be converted into $\nu_s$ ($\bar\nu_s$) 
after passing through the IR ($P_{\rm hop}\sim 0$), while most of those 
with $E_\nu>E_{0.5}$ will survive in their initial flavor states ($P_{\rm hop}\sim 1$).

The sensitivity of $P_{\rm hop}$ to $dY_e/dr$ at $Y_e\approx 1/3$ requires
special attention. In the dynamic environment of a supernova, the $k$th mass element
is characterized by its radius $r_k(t)$, temperature $T_k(t)$, density $\rho_k(t)$, and
electron fraction $Y_{e,k}(t)$ as functions of time $t$. The profile $Y_e(r,t)$ 
at a specific $t$ is obtained from the sets $\{Y_{e,k}(t)\}$ and $\{r_k(t)\}$ formed by
all mass elements. The time evolution of $Y_{e,k}(t)$ is governed by
\begin{eqnarray}\label{eq-ye}
\frac{dY_{e,k}}{dt}&=&[\lambda_{\nu_e n,k}(t)+\lambda_{e^+ n,k}(t)]Y_{n,k}(t)\nonumber\\
&&-[\lambda_{\bar\nu_e p,k}(t)+\lambda_{e^- p,k}(t)]Y_{p,k}(t),
\end{eqnarray}
where $Y_{n,k}(t)$ and $Y_{p,k}(t)$ are the neutron and proton fraction, respectively, and 
$\lambda_{\alpha\beta,k}(t)$ corresponds to the rate per target nucleon for reactions 
(\ref{eq-nuen}) and (\ref{eq-anuep}) and their reverse reactions in the mass element. 
For $T_k\gtrsim 10^{10}$~K, $Y_{n,k}\approx 1-Y_{e,k}$ and 
$Y_{p,k}\approx Y_{e,k}$. In general, $Y_{n,k}(t)$ and $Y_{p,k}(t)$ can
be followed with a nucleosynthesis network given $T_k(t)$, $\rho_k(t)$, and $Y_{e,k}(t)$, 
from which $\lambda_{e^+ n,k}(t)$ and $\lambda_{e^- p,k}(t)$ can also
be calculated. As $\lambda_{\nu_e n,k}(t)$ and $\lambda_{\bar\nu_e p,k}(t)$
used to determine $Y_{e,k}(t)$ are affected by $\nu_e$-$\nu_s$ and $\bar\nu_e$-$\bar\nu_s$
conversion, which in turn depends on $Y_e(r,t)$ obtained from the set $\{Y_{e,k}(t)\}$,
we must treat this feedback in calculating $P_{\rm hop}$.

We use the data from an $8.8\,M_\odot$ electron-capture supernova (ECSN) 
simulation~\cite{Fischer:2009af},
which features the only successful explosion in spherical symmetry with 
three-flavor Boltzmann neutrino transport \cite{Fischer:2009af,Huedepohl:2009wh}.
In this model, an early onset of the explosion occurs at time post core bounce
$t_{\rm pb}\sim 38$~ms, in qualitative agreement with
multi-dimensional simulations \cite{Janka:2007di}. We take the sets $\{r_k(t)\}$, $\{T_k(t)\}$, 
and $\{\rho_k(t)\}$ from the simulation and increase the resolution by adding $\sim$2,000 mass 
elements in the IR region so that the resonance can be resolved properly. We recalculate
each $Y_{e,k}(t)$ using Eq.~(\ref{eq-ye}) to obtain the self-consistent $Y_e(r,t)$
in the presence of ASFC. 
As initial values of $Y_{e,k}(t)$, we use the $Y_e$ profile of the simulation at
$t_{\rm pb}\approx 30$~ms when the shock has already passed through the IR region.
Subsequently, 
we use the recalculated high-resolution $Y_e$ profile to determine which 
$\nu_e$ ($\bar\nu_e$) have 
crossed the IR and compute their hopping probabilities, $P_{{\rm hop}}(E_\nu,\mu)$.
These probabilities are then multiplied by the distribution function of the $\nu_e$ 
($\bar\nu_e$), $f_{\nu_e(\bar\nu_e)}(E_\nu,\mu)$, given by the simulation to determine 
$\lambda_{\nu_e n,k}$ ($\lambda_{\bar\nu_e p,k}$) in Eq.~(\ref{eq-ye}).

\begin{figure}
\includegraphics[angle=-90,width=\linewidth]{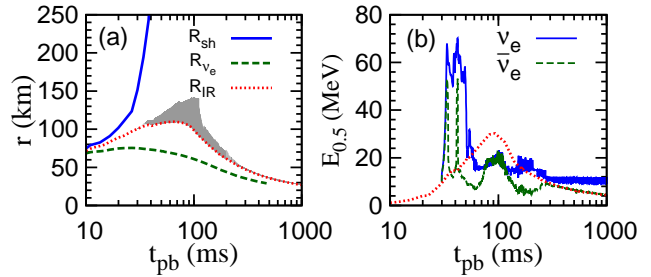}
\caption{
(Color online) (a) Evolution of $R_{\nu_e}$ (green dashed curve), $R_{\rm sh}$ (blue solid curve), 
and $R_{\rm IR}$ for the original (red dotted curve) and recalculated (shaded region) 
$Y_e$ profiles.
(b) Evolution of $E_{0.5}$ for the original (red dotted curve) and 
recalculated (blue solid and green dashed curves) $Y_e$ profiles.
All results for ASFC assume $\delta m^2=1.75$~eV$^2$ and $\sin^2{2\theta}=0.10$.
\label{fig-IR-Eh}}
\end{figure}

We first consider the results for $\delta m^2=1.75$~eV$^2$ and $\sin^2{2\theta}=0.10$.
We compare the original $Y_e$ profile at $t_{\rm pb}\sim 34$~ms with the 
one obtained by including ASFC feedback in Fig.~\ref{fig-profile}c.
ASFC results in lower $Y_e$ values and produces a plateau at $Y_e=1/3^+$ corresponding
to $r\sim 106$~km. The time evolution of $R_{\nu_e}$, shock radius 
$R_{\rm sh}$, and $R_{\rm IR}$ and $E_{0.5}$
with and without ASFC feedback is shown in Fig.~\ref{fig-IR-Eh}.
If ASFC feedback is neglected, all $\nu_e$ and $\bar\nu_e$ of different $E_\nu$ 
have approximately the same $R_{\rm IR}$. In this case,
$E_{0.5}$ and $R_{\rm IR}$ follow a similar trend: they 
increase due to the flattening of the $Y_e$ profile during the initial shock expansion, 
and decrease later due to the steepening of the $Y_e$ profile
during the protoneutron star cooling.
Inclusion of ASFC feedback significantly extends the IR region
in radius for times, $t_{\rm pb}\sim 32$--200~ms, 
due to the formation of a plateau at $Y_e= 1/3^+$.
The $Y_e$ plateau greatly reduces $|dY_e/dr|$ in the IR region and affects $\nu_e$ more than 
$\bar\nu_e$ as the latter cross the IR at $Y_e= 1/3^-$. Consequently,
$E_{0.5}$ differs for $\nu_e$ and $\bar\nu_e$ with 
$E_{0.5,\nu_e}\gtrsim E_{0.5,\bar\nu_e}$.
Compared to the case without ASFC feedback,
both $E_{0.5,\nu_e}$ and $E_{0.5,\bar\nu_e}$ rise much faster to larger values greatly 
exceeding $\langle E_\nu\rangle$ during the initial shock expansion,
i.e., most of $\nu_e$ and $\bar\nu_e$ are converted into sterile counterparts.
At later times, $E_{0.5,\bar\nu_e}$ decreases while
$E_{0.5,\nu_e}$ remains at $\sim 10$~MeV~$\sim\langle E_{\nu_e}\rangle$.

\begin{figure}
\includegraphics[angle=-90,width=\linewidth]{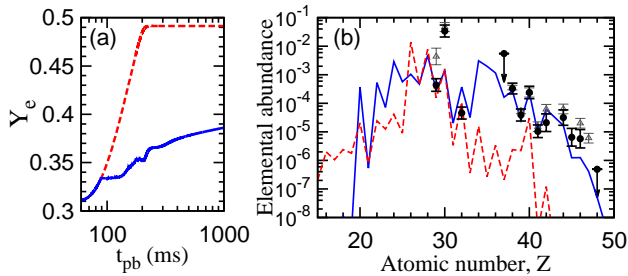}
\caption{
(Color online) Comparison of (a) $Y_e$ evolution in
an example mass element and (b) integrated nucleosynthesis 
with (blue solid curve) and without (red dashed curve) 
ASFC. Data on the metal-poor star HD 122563 normalized to
the calculated abundance of Zr ($Z=40$) with ASFC
are shown as open triangles \cite{Honda:2006kp} and filled circles \cite{Roederer:2012dr} in (b). 
\label{fig-nusyn}}
\end{figure}

The effect of the plateau at $Y_e= 1/3^+$ on $E_{0.5,\nu_e}$
enhances conversion of $\nu_e$ into $\nu_s$, which greatly decreases $Y_e$ 
at larger radii by reducing the rate of reaction~(\ref{eq-nuen}).
We show the $Y_e$ evolution for an example mass element as a function of time in 
Fig.~\ref{fig-nusyn}a with and without ASFC feedback. 
As the mass element is being ejected, it encounters the plateau of $Y_e= 1/3^+$ at 
$t_{\rm pb}\sim 90$--120~ms and 
its $Y_e$ is greatly reduced from the original supernova simulation 
values (from 0.49 to 0.37--0.39 for $t_{\rm pb}\gtrsim 200$~ms).
This reduction is mainly driven by $e^-$ capture on protons,
the inverse of reaction~(\ref{eq-nuen}), and by absorption on protons of
the surviving $\bar\nu_e$, reaction~(\ref{eq-anuep}), after most $\nu_e$
have been converted into $\nu_s$.

Similar reduction of $Y_e$ by ASFC occurs in $\sim 10^{-2}$~$M_\odot$ of 
ejecta. The integrated nucleosynthesis in this material
is shown in Fig.~\ref{fig-nusyn}b. Compared to the case without ASFC where
only elements with $Z\lesssim 30$ are produced, much heavier elements
from $Z=38$ (Sr) to $Z=48$ (Cd) are produced with ASFC and their
pattern is in broad agreement with observations of the metal-poor star 
HD 122563 \cite{Honda:2006kp,Roederer:2012dr}.
It remains to be explored if ASFC can help to overcome the difficulties
of neutrino-driven winds from more massive supernovae in producing elements 
with $Z>42$ \cite{Martinez-Pinedo:2013jna}.
ECSN differs from those models by the presence of a dynamically ejected 
neutron-rich component \cite{Wanajo:2010ig}. This material is ejected at
$t_{\rm pb}\lesssim 100$~ms and its $Y_e$ is reduced to $\sim 0.38$ by ASFC,
thereby enabling production of elements with $Z>42$.

\begin{figure}
\centering
\includegraphics[angle=-90,width=1.0\linewidth]{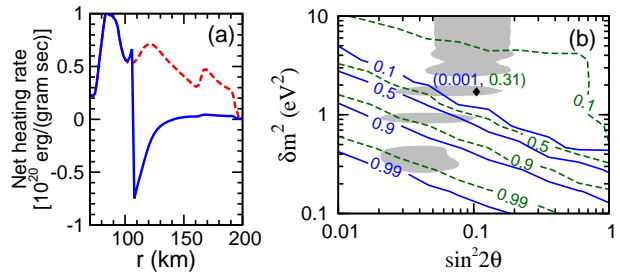}
\caption{
(Color online) (a) Comparison of the net heating rate with (blue solid curve) 
and without (red dashed curve) ASFC
as a function of radius at $t_{\rm pb}\sim 34$~ms for $\delta m^2=1.75$~eV$^2$ 
and $\sin^2{2\theta}=0.10$ [filled diamond in (b)]. 
(b) Contours of $\dot q'_{\nu_en}/\dot q_{\nu_en}$ (blue solid curves) and 
$\dot q'_{\bar\nu_ep}/\dot q_{\bar\nu_ep}$ (green dashed curves) for
$r>R_{\rm IR}$ at $t_{\rm pb}\sim 34$~ms. Numbers in parentheses
give ($\dot q'_{\nu_en}/\dot q_{\nu_en},\dot q'_{\bar\nu_ep}/\dot q_{\bar\nu_ep}$)
for the mixing parameters used in (a). 
Shaded regions give 
mixing parameters inferred at the 90\% confidence level by \cite{Kopp:2013vaa}.
\label{fig-mixing}}
\end{figure}

In Fig.~\ref{fig-mixing}a we compare the net heating rate 
(neutrino heating minus matter cooling)
with and without ASFC as a function of radius at
$t_{\rm pb}\sim 34$~ms. 
It can be seen that ASFC drastically turns the region of net heating at 
$r\gtrsim 110$~km into one of net cooling. So far we have focused on the results for
$\delta m^2=1.75$~eV$^2$ and $\sin^2{2\theta}=0.10$. In view of the potential 
implications of ASFC for supernova explosion, we examine a wide range of 
mixing parameters. In Fig.~\ref{fig-mixing}b we show contours of 
$\dot q'_{\nu_en}/\dot q_{\nu_en}$ and $\dot q'_{\bar\nu_ep}/\dot q_{\bar\nu_ep}$
for $r>R_{\rm IR}$ at $t_{\rm pb}\sim 34$~ms in the ($\sin^22\theta,\delta m^2$) space, 
where $\dot q'_{\nu_en}$ and $\dot q'_{\bar\nu_ep}$ are the heating rates for 
reactions~(\ref{eq-nuen}) and (\ref{eq-anuep}) with ASFC, respectively, and
the unprimed counterparts are for the case without ASFC.
The filled diamond in Fig.~\ref{fig-mixing}b
represents the mixing parameters used above and the shaded 
regions give those inferred from reactor neutrino experiments
at the 90\% confidence level \cite{Kopp:2013vaa}. 
Except for the two regions with the lowest $\delta m^2$, all other inferred 
parameters for ASFC might have a large negative impact on the explosion 
of the $8.8\,M_\odot$ ECSN.

We have shown that the existence of sterile neutrinos with parameters 
inferred from reactor neutrino experiments produces substantial ASFC of the 
$\nu_e$-$\nu_s$ and $\bar\nu_e$-$\bar\nu_s$ types near the core of
an $8.8\,M_\odot$ ECSN. 
As a result of ASFC feedback, a $Y_e$ plateau is formed in the resonance region 
where $Y_e\approx 1/3$.
This further enhances conversion of $\nu_e$ 
into $\nu_s$, thereby reducing $Y_e$ at larger radii. For the inferred 
best-fit parameters, nuclei with $Z>40$ are produced in a total
$\sim 10^{-2}\,M_\odot$ of supernova ejecta with a pattern
in broad agreement with metal-poor star observations. 
Without ASFC, only nuclei with $Z\lesssim 30$
are produced. However, for a wide range of mixing parameters, 
the neutrino heating rates are strongly suppressed by ASFC
at times when such heating is important for energizing the shock. 
A caveat of our treatment is that suppression of neutrino heating by ASFC
would likely change the dynamic and thermodynamic conditions.
Thus, the
exact effects of ASFC on supernova explosion and nucleosynthesis
remain to be studied by implementing ASFC in the simulations self-consistently.
These studies should also be extended to supernova models for more 
massive progenitors. Our results suggest that such 
studies can strongly constrain the mixing parameters for ASFC in 
combination with neutrino experiments and cosmological considerations.
In the future we will examine the effects of ASFC in supernovae along with other 
flavor conversion processes and determine the impact on neutrino signals in 
terrestrial detectors. These studies along with self-consistent 
treatment of neutrino flavor transformation in supernovae can not only 
provide unique probes of neutrino mixing, but may also
help understanding supernova explosion and nucleosynthesis.

M.-R.W. is supported by the Alexander von Humboldt Foundation.
T.F. acknowledges support from the Narodowe Centrum Nauki (NCN) within
the "Maestro" program under contract No. DEC-2011/02/A/ST2/00306.
L.H. and G.M.P. are partly supported by the Deutsche Forschungsgemeinschaft
through contract SFB 634, the Helmholtz International Center for FAIR
within the framework of the LOEWE program launched by the state of
Hesse and the Helmholtz Association through the Nuclear Astrophysics
Virtual Institute (VH-VI-417). Y.-Z.Q. is partly supported by the US DOE (DE-FG02-87ER40328).
We gratefully thank Hans-Thomas Janka, Irene Tamborra, and two anonymous reviewers for
helpful comments and suggestions.


%

\end{document}